\documentclass[conference,twoside,table]{IEEEtran}
\IEEEoverridecommandlockouts
\usepackage{cite}
\usepackage{amsmath,amssymb,amsfonts}
\usepackage{algorithmic}
\usepackage{graphicx}
\usepackage{textcomp}
\usepackage{xcolor}
\usepackage{booktabs}
\usepackage{todonotes}
\usepackage[hyphens]{url}

\begin{document}

\title{A traceability and auditing framework for electronic equipment reverse logistics based on blockchain: the case of mobile phones}

\author{\IEEEauthorblockN{Thomas K. Dasaklis}
\IEEEauthorblockA{\textit{Department of Informatics} \\
\textit{University of Piraeus}\\
Piraeus, Greece \\
dasaklis@unipi.gr}
\and
\IEEEauthorblockN{Fran Casino}
\IEEEauthorblockA{\textit{Department of Informatics} \\
\textit{University of Piraeus}\\
Piraeus, Greece \\
francasino@unipi.gr}
\and
\IEEEauthorblockN{Constantinos Patsakis}
\IEEEauthorblockA{\textit{Department of Informatics} \\
\textit{University of Piraeus}\\
Piraeus, Greece \\
kpatsak@unipi.gr}
}

\maketitle

\begin{abstract}

Human beings produce electronic waste (e-waste) at an unprecedented pace. Mobile phones and other inter-connected smart devices make a significant contribution to the generation of e-waste. Reverse logistics (RL) activities play an essential role in managing mobile phones during their end-of-life. However, remanufacturing and/or refurbishing of mobile phones might prove difficult not only from an operational point of view but also from a data management and privacy perspective (due to privacy-related regulatory frameworks like the EU General Data Protection Regulation directive). In this paper, we propose a distributed trustless and secure framework for electronic equipment RL activities based on blockchain technology. We consider the remanufacturing/refurbishing recovery option for mobile phones, and we develop an autonomous and effective back-end data sharing architecture based on smart contracts/blockchain technology for keeping track of all the remanufacturing/refurbishing processes. For demonstrating the applicability of our approach, we develop a functional set of smart contracts and a local private blockchain. The benefits of our framework are further discussed, along with fruitful areas for future research.

\end{abstract}


\section{Introduction}

Human beings produce electronic waste (e-waste) at an unprecedented pace. Until 2016, the world generated 44.7 million metric tonnes of e-waste, and only 20\% of this tonnage found its way through proper recycling channels \cite{balde2017global}. Although e-waste may represent only 2\% of solid waste streams, yet it can represent 70\% of the hazardous waste that ends up in dumpsites and landfills \cite{WorldEconomicForum2019}. Inappropriate disposal is a huge problem from various viewpoints since e-waste contains a multitude of toxic substances dangerous to both humans and the environment. Typical poisonous substances found in e-waste include but are not limited to lead, cadmium, lead oxide, mercury, chromium, brominated flame retardants, and polyvinyl chloride. Beyond environmental and health-related concerns, e-waste presents also a significant economic loss and tremendous loss of scarce and valuable raw materials. In particular, the remaining value of all raw materials prevalent in e-waste is estimated at approximately 55 Billion Euros in 2016. It is worth noting that this amount represents the 2016 Gross Domestic Product of most countries in the world \cite{balde2017global}. Therefore, apart from major environmental concerns, e-waste presents significant economic opportunities, and e-waste management should be an integral part of the circular economy concept.

% Please add the following required packages to your document preamble:
% \usepackage{graphicx}
% \usepackage[table,xcdraw]{xcolor}
% If you use beamer only pass "xcolor=table" option, i.e. \documentclass[xcolor=table]{beamer}
\begin{table*}[th]
\caption{Barriers for implementing RL activities. Adapted from \cite{Agrawal201576,Govindan2018318,Prakash2015,Sirisawat2018}}
\label{fig:Tab1}
\resizebox{\textwidth}{!}{%
\begin{tabular}{|l|l|}
\hline
\rowcolor[HTML]{C0C0C0} 
\multicolumn{1}{|c|}{\cellcolor[HTML]{C0C0C0}\textbf{Barriers}} & \multicolumn{1}{c|}{\cellcolor[HTML]{C0C0C0}\textbf{Description}} \\ \hline
Technological & Lack of  proper IT systems to support RL activities \\ \hline
Legal & \begin{tabular}[c]{@{}l@{}}Lack of common regulatory frameworks \\ Lack of enforced law and relevant regulations\end{tabular} \\ \hline
Financial/Economical & \begin{tabular}[c]{@{}l@{}}Increased cost associated to RL activities (hazardous material etc)\\ Lack of investment for RL activities\\ Lack of economies of scale\\ Marginal profit due to customer perception of refurbished and/or remanufactured products.\end{tabular} \\ \hline
Product & \begin{tabular}[c]{@{}l@{}}Modularity\\ Uncertainty of reverse product flows\end{tabular} \\ \hline
Available infrastructure & Lack of appropriate infrastructure to support RL activities (storage facilities, vehicles etc). \\ \hline
Managerial/ Organizational & \begin{tabular}[c]{@{}l@{}}Lack of commitment from top management\\ Lack of proper organizational structures to support RL activities\\ Integration difficulties of RL partners.\\ Collaboration and coordination challenges of disparate RL members\end{tabular} \\ \hline
Market-oriented & \begin{tabular}[c]{@{}l@{}}Marketing of refurbished and/or remanufactured products\\ Customer perception of refurbished and/or remanufactured products\end{tabular} \\ \hline
Policy and governance & \begin{tabular}[c]{@{}l@{}}Lack of standards for RL operations\\ Lack of waste management practices\\ Geographical challenges for adopting ``extended producer responsibility'' approaches.\end{tabular} \\ \hline
\end{tabular}%
}
\end{table*}

Mobile phones and other inter-connected smart devices make a significant contribution to the generation of e-waste, and end-of-life treatment of these devices is of paramount importance. Like e-equipment, mobile phones contain toxic substances that may cause harm if disposed of inappropriately. A significant driver behind the increased amount of e-waste generated by mobile phones may relate to consumer habits and the growing ``throwaway society''. For example, in the United States mobile phones are used on average 20 months before being replaced, although they can function properly for a more extended period of time \cite{balde2017global}. The increased replacement rate of smartphones may also relate to a) consumers' social and interpersonal environment (influencing their decisions on when to change a mobile phone device), b) emerging technologies and performance improvements, c) greater variety in applications and faster networking and speeds of newer mobile phones.

Reverse logistics (RL) activities play an essential role in managing e-waste and further prolonging the life-cycle of e-equipment. Various definitions exist in the literature relevant to RL \cite{Kazemi20194937,Hall2013}. For example, the American Reverse Logistics Executive Council defines RL as ``\textit{The process of planning, implementing, and controlling the efficient, cost-effective flow of raw materials, in-process inventory, finished goods and related information from the point of consumption to the point of origin for the purpose of recapturing value or proper disposal}'' \cite{Govindan2017371}. A more practical definition of RL may include the end-of-life collection of products and their repairing, disassembling, remanufacturing, recycling, and disposing of them to recapture value or ensure proper disposal \cite{Prajapati2019503}. Therefore, RL provides the means for minimizing the waste generated in the supply chain (SC) by recovering the remaining amount of used materials and resources and producing new value-added products in an environmentally sustainable manner \cite{Jayasinghe2019}. 

Despite the significant benefits of RL activities to both the environment and the economy, several barriers hinder the establishment of robust RL mechanisms. As shown in Table \ref{fig:Tab1}, a significant barrier for implementing RL activities is the lack of common regulatory frameworks and relevant policies. Of particular interest may be the geographical challenges for adopting ``\textit{extended producer responsibility}'' approaches across geographically dispersed logistics networks, especially in the case of e-equipment. Financial/economic issues like the increased cost associated with RL activities and lack of investments also act as barriers for implementing sound RL activities. Lack of proper organizational structures to support RL activities and collaboration and coordination challenges of disparate RL members also hinder the development of RL practices. Other barriers may relate to the lack of appropriate infrastructure to support RL activities; like storage facilities, vehicles, etc., as well as the inherent uncertainty of the products' reverse flows.

\subsection{Motivation and contribution}
\label{sec:motivation}
Apart from the RL barriers mentioned above, remanufacturing and/or refurbishing of e-equipment (laptops, tablets, mobile phones etc) might prove difficult, especially from a data management and privacy perspective. In fact, user data pose a significant challenge for a manufacturer/third party when repairing/refurbishing a mobile phone device. Privacy-related regulatory frameworks (like the EU General Data Protection Regulation directive) put additional pressure on refurbishing parties for regulatory compliance and for effectively managing mobile devices containing user data with varying levels of sensitivity. Other important issues raised may include, for example, which party assumes liability for the data stored in returned mobile phones and which stakeholder eventually retains ownership of the data and the overall refurbishing processes. Therefore, it is evident that mobile phones RL activities call for new approaches, which incorporate strong process integrity guarantees and forensics-by-design concepts.

The paper addresses the aforementioned gaps in the literature by providing a distributed trustless and secure architecture for e-equipment RL activities based on blockchain technology. For demonstrating the potential of the proposed architecture to support a diversity of capabilities and requirements we consider the remanufacturing/refurbishing recovery option for mobile phones. We develop an autonomous and effective back-end data sharing architecture based on smart contracts/blockchain technology for keeping track of all the remanufacturing/refurbishing processes taking place by all the actors involved. This forensics-by-design framework safeguards the chain-of-custody for all the remanufacturing/refurbishing activities taking place, thus building trust in RL operations not only from a data privacy perspective but from a market viewpoint as well, thus increasing customers trust and appreciation to refurbished phones. It should be stated that recycling processes have been excluded from the analysis since recycling is a relatively straightforward process in terms of handling sensitive data etc (as opposed to remanufacturing/refurbishing in which special attention should be paid to remaining customer data). To the best of our knowledge, this is the fist blockchain-enabled approach in RL and remanufacturing/refurbishing of used mobile phones.

% \todo[inline,color=cyan]{While recycling procedures can be also audited by our framework to enable, for instance, some statistics or provide efficacy flows, we focus on the other two end-of-life cycles (i.e.remanufacturing/refurbishing) due to their challenges in terms such as GDPR compliance, trust and economic impact, leaving further discussion of recycling for future work.} 

\subsection{Paper structure}

The remainder of the paper is organized as follows. In Section \ref{sec:Sustainable_SCM}, we provide an overview of Industry 4.0 applications for SC sustainability with a special focus on blockchain technology. In Section \ref{sec:Literature}, we briefly review the available literature relevant to mobile phones and RL activities. In Section \ref{sec:Framework}, the proposed blockchain-enabled framework is presented, and we detail our proof of concept implementation. Then, in Section \ref{sec:Discussion}, we discuss the various benefits of the proposed framework, along with fruitful areas for future research. The paper ends in Section \ref{sec:Conclusions} with some concluding remarks.

\section{Industry 4.0 and supply chain sustainability}
\label{sec:Sustainable_SCM}

The Fourth Industrial Revolution is characterized by the convergence of various technologies, such as the Internet of Things (IoT) and blockchain, which lead to the complete digitization of SC networks. There seems to be a unique opportunity to harness the Fourth Industrial Revolution for improving the sustainability of SC networks \cite{Stock2016}. For instance, prospects for sustainable SC practices exist by designing products for longevity, repair, and recycling. Such practices transform the overall value proposition, SC management, relation with the customers and financial justification of business models \cite{Man2017}. Besides, Industry 4.0 initiatives can significantly influence business systems via transforming how the products are designed, produced, delivered, and discarded \cite{Luthra2018} or for improving environmental performance that can open up new market segments to companies until previously unexplored \cite{Garcia-Muina2019}. Advanced and digital manufacturing technologies have the potential to unlock the circularity of resources within SC networks and assist in sustainable operations management decision-making \cite{LopesdeSousaJabbour2018}.

% \subsection{Sustainable supply chain management and blockchain technology}

Blockchain technology, a foundational technology of the Fourth Industrial Revolution, is expected to play a crucial role in all aspects of SC management \cite{CASINO201955}. Blockchain, in the simplest of terms, is a distributed and incorruptible digital ledger, that enables the transfer of a range of assets among non-trusted parties in a secure and encrypted way without third-party intermediaries \cite{CASINO201955}. Additional functionalities may be achieved in blockchain technology with the incorporation of two features: smart contracts (SmCs) and tokens. SmCs are agreements in the form of computer programs among mutually distrusting participants that may be executed in distributed environments and are automatically enforced by the consensus mechanism of the blockchain (without relying on a trusted third party) \cite{christidis2016blockchains}. Tokens are digital entities that may be used as a digital representation of physical assets, usually in decentralized applications. 

Blockchain technology is expected to play an essential role in all three pillars of SC sustainability (economic, environmental, and social). For instance, by providing trust in distributed environments, blockchain has the potential to offer full transparency and visibility within the SC \cite{Nikolakis2018,Saberi2019} and further enhance SC traceability processes \cite{Dasaklis_COINS}. Blockchain technology may also build trust among disparate SC members and further improve coordination and information sharing \cite{DiVaio2019}. Various applications exist in the literature in which blockchain technology is used as a possible candidate for making local, green energy production \cite{Lund2019} or as a green asset management platform to support low-carbon emission technologies \cite{Wu2018}. It is worth noting that blockchain-enabled SC approaches coupled with IoT applications, could improve the communication and the selective export of SC traceability data, enabling additional benefits to the logistics sector for data management and analytics \cite{banafa2017iot}.

\section{Literature review}
\label{sec:Literature}

Several approaches and theoretical frameworks exist in the literature for RL activities \cite{Agrawal201576,Hall2013,Islam2018,Pokharel2009175}. Regarding RL activities for mobile phones, two main streams of research may be identified. The first stream relates to qualitative studies for mobile phones RL. For example, in \cite{Chan2006}, a conceptual model for mobile phones RL is presented. The model takes into account internal and external factors affecting material flows and the RL activities. Documentation studies of country-specific approaches for mobile phones RL activities also exist in the literature. For example, in \cite{Ponce-Cueto2011} a characterization of the different actors involved in the mobile phones RL systems in  Spain is provided. In \cite{Maheswari2017} a quattro helix model (manufactures, government, takeback operators, and societies) is presented for building robust RL programs for mobile phones whereas in \cite{Maheswari2019} a value chain analysis approach is adopted for defining the interrelationships, expectations, and values of the various stakeholders involved in mobile phones RL activities.

Some quantitative RL approaches have also been developed in the literature. In \cite{Fernandez2006}, a model is developed for optimally designing the RL network for recovering mobile phones. The main objective of the model is the minimization of the collection, transportation, management, and storage costs. An integer linear programming model for the network design of a multi-echelon, multi-product RL network is presented in \cite{John2018}. The model incorporates products' structure as well as various cost elements (transportation, fixed facility, processing, collection, and disposal costs). Finally, in \cite{Jayant2014} a decision support system is presented for the selection and evaluation of different 3PL RL service providers to provide mobile phone RL operations. The Analytical Hierarchy Process (AHP) and Technique for Order Preference by Similarity to Ideal Solution (TOPSIS) methods are applied for the development of the decision support system. As previously stated in Section \ref{sec:motivation} it is worth noting that we did not manage to retrieve any study from the literature relevant to blockchain technology and RL activities.

\section{The proposed reverse logistics framework}
\label{sec:Framework}

RL activities and end-of-life management of e-equipment entail various operational challenges and a multitude of processes and stakeholders/players. In this section, we describe our blockchain-based RL framework for mobile phones and we further elaborate on the various RL players involved and their role and the RL processes taking place.
% \todo[inline, color=cyan]{FC THE IDEA HERE IS TO STATE WHAT WE WANT TO SOLVE- As previously stated, the end-of-life cycle of an electronic device entails many challenges, actors, processes and stakeholders involved. In this section, we describe our blockchain-based RL framework for the case of mobile phones to provide a comprehensive flow of their interactions and the features/benefits enabled by it. First, we present the details of the RL flow, and then we define the main actors/processes of our framework and their interrelationships.}

\begin{figure}[th]
    \centering    
    \includegraphics[width=\columnwidth]{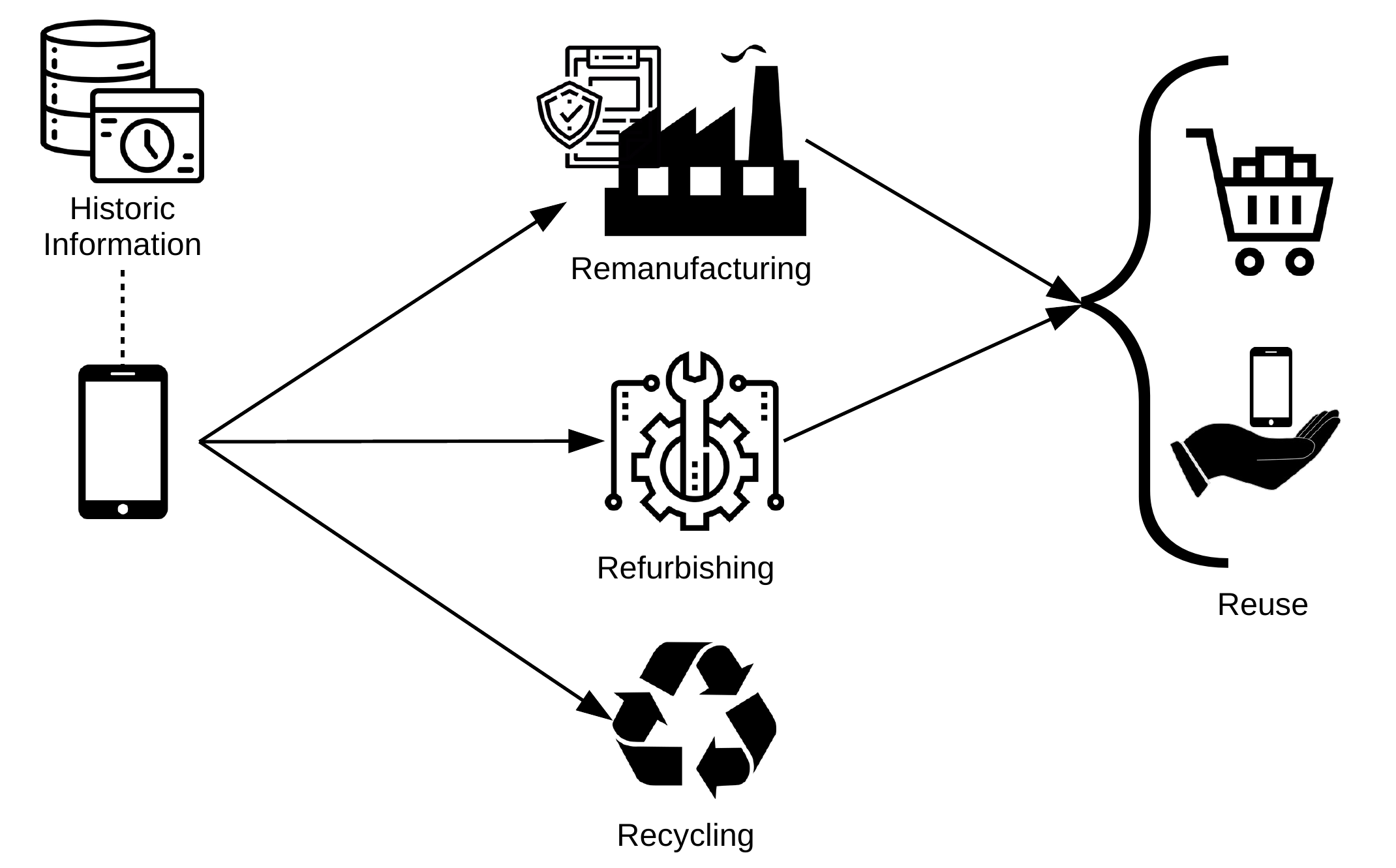}
    \caption{Overview of the mobile phones reverse logistics flow. }
    \label{fig:flow}
\end{figure}

RL activities for mobile phones may assist either to a) their safe end-of-life treatment (recycling) and/or proper disposal or b) their refurbishment/manufacturing and their subsequent sale to secondary or emerging foreign markets. The RL flow for mobile phones can be seen in Figure \ref{fig:flow}. Depending on the conditions of the phone and the interested parties, the RL activities for the mobile phone may be: remanufacturing, refurbishing or recycling. It should be noted that refurbished phones and used phones are not the same. Used phones, once having been wiped of information, are typically sold as-is. A re-manufactured mobile device is practically a new device meeting the original manufacturing standards, whereas a refurbished phone by a third party may be meeting lower quality standards. 

Note that all existing information about the mobile phone will be stored and remain accessible regardless of the path followed. In this paper, we focus on the first two paths (remanufacturing and/or refurbishing), leaving the recycling path and its challenges for future research (like, for instance, conforming with environmental regulations for recycling process etc). Therefore, whether a phone has been remanufactured or refurbished, it will be either sold in a secondary market or donated.

\subsection{Key players and reverse logistics operations}\label{sec:players}
In what follows, we describe the characteristics of the main actors/resources of the framework, namely the products, the stakeholders, and the processes.

\begin{figure}[th]
    \centering    
    \includegraphics[width=\columnwidth]{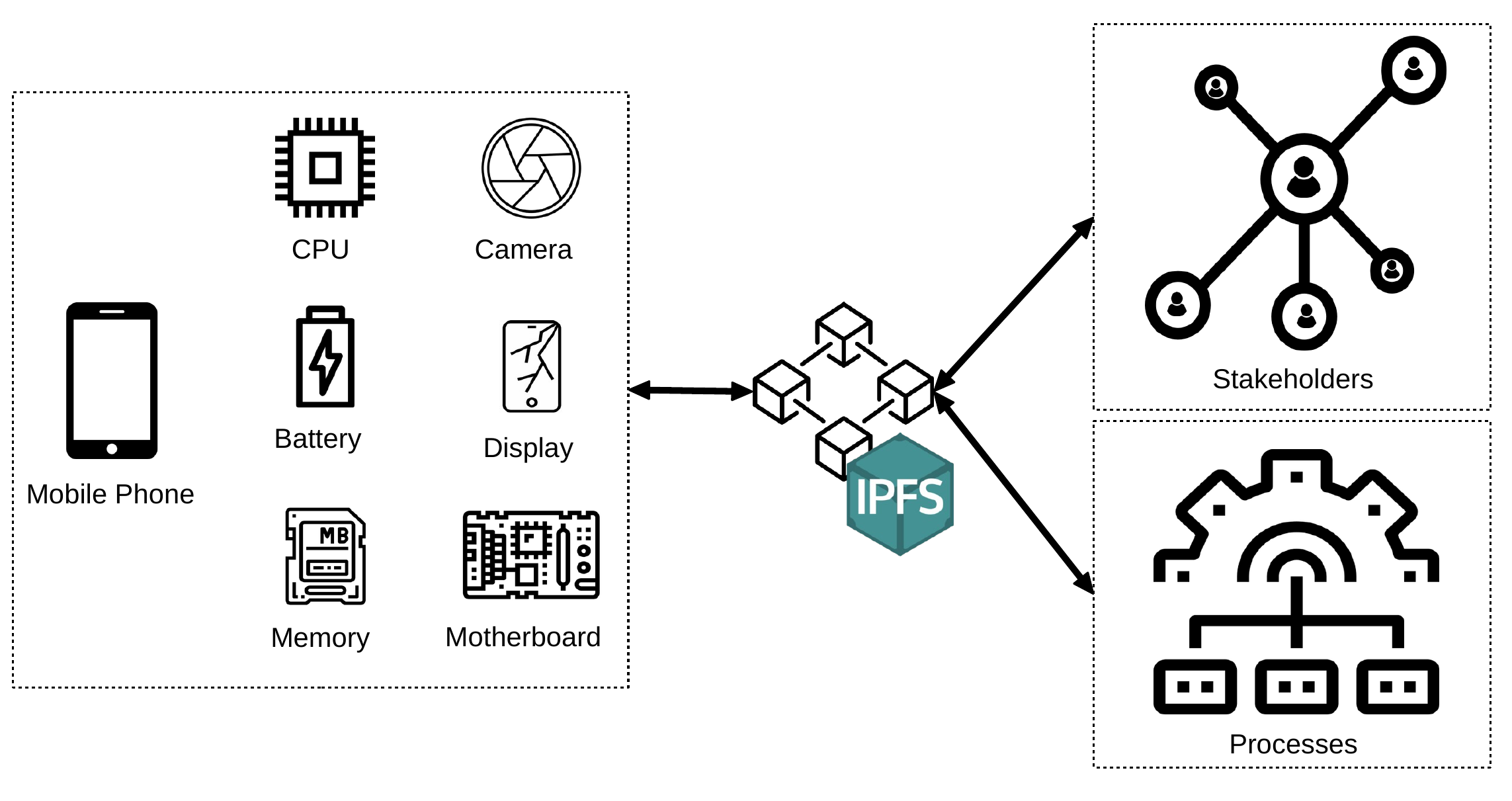}
    \caption{Interaction of the various mobile phones parts, stakeholders and refurbishing process with the blockchain and IPFS.}
    \label{fig:arch}
\end{figure}

\noindent\textbf{Product:} The main components of a mobile phone are the CPU, camera, battery, display, internal memory and the motherboard, as depicted in Figure \ref{fig:arch}. Each of these components, as well as the whole device, have their corresponding serial numbers and feature information that has to be considered to evaluate their quality and origin.

\noindent\textbf{Processes:} As previously seen in Figure \ref{fig:flow}, we can identify the main path that a mobile device can follow once entering the RL channel. In general, no standard industry definition exists for either re-manufacturing or refurbishing, and usually, the two terms are used interchangeably. Re-manufacturing entails the repairing of a mobile device by the very same manufacturer who built the phone in the first place. A manufacturer refurbished phone will look ``like-new'', and it will be resold for a slightly lower price. On the other hand, refurbishing may refer to a range of processes applied by third parties. These processes include device inspection, analysis of the device's physical condition, wiping off any information and restoring it to the original factory settings, testing for functional performance, customization's removal (either hardware or software), and repairing of damaged parts. The aforementioned processes may be seen in Figure \ref{fig:scsinterrelation}. Obviously, additional activities and more detailed procedures can be used and registered properly in our blockchain-enabled system.

\noindent\textbf{Stakeholders:} The stakeholders and their roles in mobile phone RL can be diverse. In our mobile RL approach, we consider the following players/stakeholders: customers, the retailers, manufacturers, 3PL companies and the government (Figure \ref{fig:scsinterrelation}). A mobile phone may be refurbished by the original vendor/retailer or some other third party. The role played by each stakeholder is different, for example, government acts as auditors and may check for regulatory compliance, the customers start the RL flow by bringing back their used devices to pre-defined collecting locations managed by retailers etc. 

The inherent features of blockchain technology enable the above mentioned RL stakeholders to maintain a safe, permanent, and tamper-proof digital record of transactions, without the interference of a central trusted authority. Moreover, due to its distributed nature, blockchain may simultaneously assure the availability and resilience of the RL management system efficiently. This is of great importance because strong process integrity guarantees are provided for all the refurbishing/remanufacturing activities taking place by a third party (refurbisher).

Since blockchain technology is not suitable for storing a large amount of data, we apply an off-chain storage approach (such as the InterPlanetary File System-IPFS\footnote{https://ipfs.io/}) to enable scalability and integrity for storing and retrieving vital records of the overall RL activities (as seen in Figure \ref{fig:arch}). In particular, each stakeholder stores locally (using decentralized storage like IPFS) all the relevant RL information and critical characteristics of the overall RL operations in a Table of Content (TOC) approach. Then, the hashes of all the individual records (from all the TOCs) are stored on the blockchain. Therefore, instead of storing data directly on the blockchain, we store hashes of RL data on the blockchain. Based on this approach, we manage to efficiently retrieve information for tracking the various RL activities by only using the corresponding TOC from each stakeholder.

\subsection{Experimental Setup}
\label{sec:experiments}

\begin{figure}[th]
    \centering    
    \includegraphics[width=\columnwidth]{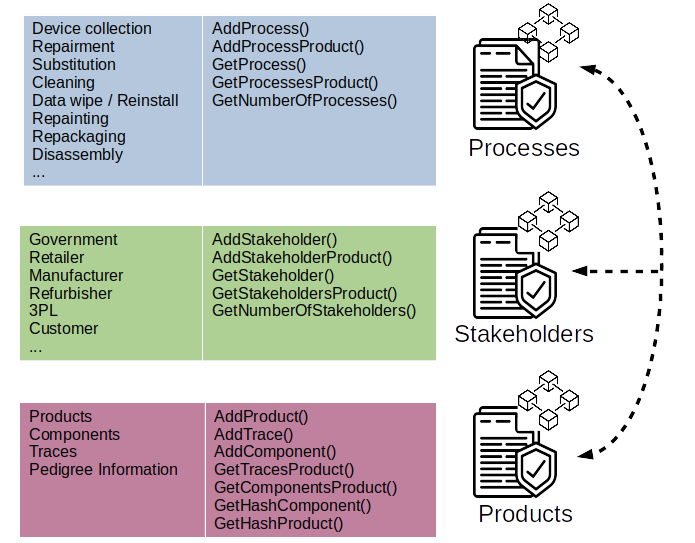}
    \caption{From left to right, examples of each key player as defined in Section \ref{sec:players}, most relevant functions implemented, and smart contracts interrelations.}
    \label{fig:scsinterrelation}
\end{figure}

A local Ethereum blockchain was created using \texttt{node}\footnote{\url{https://nodejs.org/}} and \texttt{ganache-cli}\footnote{\url{https://github.com/trufflesuite/ganache-cli}}, and \texttt{truffle}\footnote{\url{http://truffleframework.com}} was used to compile and deploy the SmCs. We assume that the involved actors are registered in the blockchain, and they have a valid address with their respective pair of public/private keys. We developed three SmCs (i.e. stakeholders, processes, and products) that interact with each other. An overview of their main characteristics and interrelations is depicted in Figure \ref{fig:scsinterrelation} and the corresponding code and implementation are available in GitHub\footnote{\url{https://github.com/francasino/Reverse_logistics_mobile}}. The SmCs implement access control restrictions (e.g., using the \textit{require} clause in \texttt{solidity}) and permissionless functions to retrieve statistics and hashes pointing to extended information of each stakeholder, process and product. The tests performed in the local blockchain showed that the average transaction time is in the magnitude of milliseconds (e.g., contract deployment and constructor), enabling real-time RL traceability.

\section{Discussion}
\label{sec:Discussion}

In this paper we have proposed a new blockchain-based framework for RL traceability and auditability for e-equipment. For showing that the proposed architecture is versatile enough to accommodate multiple use case scenarios we have created a digital representation of all the materials, sub-assemblies and intermediate assemblies of a mobile phone as described in Section \ref{sec:Framework}. To achieve our objectives, we have used blockchain and SmCs as the building blocks of our framework. In particular, we have used blockchain as a distributed tamper-proof chain-of-custody mechanism and SmCs as an automation mechanism for managing stakeholders and processes.

 \begin{table}[ht]
 \centering
   \rowcolors{2}{gray!25}{white}
   \scriptsize
   \caption{Main benefits of the proposed blockchain-based RL framework.}
  \begin{tabular}{p{.2\columnwidth}p{.7\columnwidth}}
     \toprule
   \textbf{Challenge} &  \textbf{Blockchain-based RL}  \\
   \midrule
Trust&  Increased trust due to the forensics-by-design nature of blockchain.\\
Environmental Impact&  A more robust and efficient framework for prolonging the life-cycle of e-equipment, thus reducing its environmental impact.\\
Privacy and Security & Enforcing some certifications to avoid security issues and man-in-the-middle attacks such as, e.g. malware installation or backdoored hardware. \\
Traceability & Individual tracking of all refurbished mobile phones and their corresponding parts. \\
Governance & With the use of information hashes and tamper-proof audit logs, we can provide end-to-end product tracking in a verifiable way. \\
Auditability & End-to-end verifiable content and procedures, which can be standardised to enable certifications. \\
Efficiency & Operational gains and less paperwork for handling RL processes.  \\
     \bottomrule
   \end{tabular}
   \label{tab:properties}
 \end{table}

The proposed framework offers several benefits in terms of improved RL process management, security, and resilience. The decentralized and secure nature of blockchain safeguards the accuracy, trustworthiness, timeliness, and usability of the exchanged RL info among the various stakeholders. Also, the proposed framework, since it is based on SmCs, offers complete automation to collecting, handling, and analyzing traceability-related data and events for the various RL operations and processes. In addition, the framework satisfies the need for individually tracking each refurbished device (as discrete item) as opposed to forward logistics where traceability is performed by pallet or case. Moreover, our approach delivers other qualitative benefits that foster its adoption, such as the ones described in Table \ref{tab:properties}. More concretely, we enable better auditability (in terms of scrutiny by external stakeholders), cost reduction, and increased trust. In this regard, we also identify benefits such as a lower overhead for handling transactions and improved order fulfilment. Checking for regulatory compliance by external stakeholders is further enhanced by the usage of SmCs that improve the creation of highly robust audit trails. Regarding visibility, the framework improves and strengthens the overall RL by making data readily available to all stakeholders. Moreover, the framework presents other benefits, such as efficient storage management, verifiability, and reduced interaction and communication between stakeholders, features that are translated into a notable cost reduction. Last but not least, the proposed framework offers new opportunities for cyber and physical systems to interact with each other and share information securely. Therefore, our framework can be easily adapted and extended to other electronic equipment RL activities, enabling similar procedures to other e-devices such as tablets, laptops, computers, TVs etc.

Although the proposed traceability mechanism presents several benefits to RL implementation, some limitations should be kept in mind. 
It should be noted that, despite the hype, blockchain is an emerging technology with several limitations and technical functionalities that are not well understood across industry and academia. Among them, scalability is one of the main challenges to solve. Scalability issues mainly stem from the time required to confirm/verify transactions, which may take several minutes for popular cryptocurrencies. Therefore, private permissioned blockchains need to be used, so that scalability issues of public blockchains are overcome. In addition, in our proposed framework, we have focused on theoretical experimentation of blockchain solution without strategically evaluating its potential to create value in real-life RL operations. Therefore, a real-life scenario would provide concrete evidence on the suitability of blockchain technology in the RL sector to create value and would further assess several of its technological limitations (scalability and performance, off-chain storage efficiency, applicability in businesses scenario in terms of costs and benefits, etc.). 

Last but no least, permanent traceability records on the blockchain creates an undeniable (and potentially unavoidable) transparency of the RL processes, which can act as a double-edged sword, particularly for managing the sensitive data contained in the returned mobile devices. The inherent security of blockchain and proper permission management (paired with off-chain storage and encrypted transactions data) minimizes the risk of data disclosure and monetary losses \cite{Atzei2017ATrust,politou2018forgetting}, yet this immutability may have a negative impact, should data not be managed and stored properly \cite{8883080}. Therefore, the inherent features of blockchain enable a myriad of features and opportunities that have to be managed carefully. 

Future work, therefore, will focus on the implementation of the proposed framework in a real-life application as well as on the use of blockchain tokens \cite{Westerkamp2019,dasaklistokens} for establishing a more detailed RL framework with advanced granularity such as the one proposed in \cite{Dasaklis_COINS}. Moreover, mechanisms to ensure proper data management \cite{bernabe2019privacy,8883080} and their impact to the overall RL framework will be studied.

\section{Conclusions}
\label{sec:Conclusions}

In this paper, we have presented an innovative traceability and auditability framework for RL activities of e-equipment, with a special focus on mobile phone devices. Based on blockchain technology and its intrinsic characteristics (immutability, forensics-by-design nature, process integrity guarantees), we have tried to tackle emerging issues in mobile phones RL activities like safeguarding the chain-of-custody for all the remanufacturing/refurbishing activities taking place with a particular focus on managing retained user sensitive data. Also, we have provided a functional implementation through the use of a local private blockchain and various SmCs. Several benefits of the framework mentioned above have been presented and thoroughly discussed.

\section*{Acknowledgment}
This work was supported by the European Commission under the Horizon 2020 Programme (H2020), as part of the project \textit{LOCARD} (Grant Agreement no. 832735).

The content of this article does not reflect the official opinion of the European Union. Responsibility for the information and views expressed therein lies entirely with the authors.

\bibliographystyle{IEEEtran}
\bibliography{IEEEabrv,refs.bib}
\end{document}